\documentclass[11pt,epsf]{article}
\usepackage{graphicx}

\catcode`\@=11
\long\def\@makefntext#1{
\protect\noindent \hbox to 3.2pt {\hskip-.9pt  
$^{{\eightrm\@thefnmark}}$\hfil}#1\hfill}		

\def\@makefnmark{\hbox to 0pt{$^{\@thefnmark}$\hss}}	
	
\def\ps@myheadings{\let\@mkboth\@gobbletwo
\def\@oddhead{\hbox{}
\rightmark\hfil\eightrm\thepage}   
\def\@oddfoot{}\def\@evenhead{\eightrm\thepage\hfil
\leftmark\hbox{}}\def\@evenfoot{}
\def\sectionmark##1{}\def\subsectionmark##1{}}

\oddsidemargin=\evensidemargin
\newcounter{sectionc}\newcounter{subsectionc}\newcounter{subsubsectionc}
\renewcommand{\section}[1] {\vspace{12pt}\addtocounter{sectionc}{1} 
\setcounter{subsectionc}{0}\setcounter{subsubsectionc}{0}\noindent 
	{\tenbf\thesectionc. #1}\par\vspace{5pt}}
\renewcommand{\subsection}[1] {\vspace{12pt}\addtocounter{subsectionc}{1} 
	\setcounter{subsubsectionc}{0}\noindent 
	{\bf\thesectionc.\thesubsectionc. {\kern1pt \bfit #1}}\par\vspace{5pt}}
\renewcommand{\subsubsection}[1] {\vspace{12pt}\addtocounter{subsubsectionc}{1}
	\noindent{\tenrm\thesectionc.\thesubsectionc.\thesubsubsectionc.
	{\kern1pt \tenit #1}}\par\vspace{5pt}}

\newcommand{\textlineskip}{\baselineskip=13pt}

\def\eightcirc{
\begin{picture}(0,0)
\put(4.4,1.8){\circle{6.5}}
\end{picture}}
\def\eightcopyright{\eightcirc\kern2.7pt\hbox{\eightrm c}} 

\def\abstracts#1#2#3{{
	\centering{\begin{minipage}{4.5in}\baselineskip=10pt\footnotesize
	\parindent=0pt #1\par 
	\parindent=15pt #2\par
	\parindent=15pt #3
	\end{minipage}}\par}} 

\renewenvironment{thebibliography}[1]
	{\frenchspacing
	 \ninerm\baselineskip=11pt
	 \begin{list}{\arabic{enumi}.}
        {\usecounter{enumi}\setlength{\parsep}{0pt}     
	 \setlength{\leftmargin 12.7pt}{\rightmargin 0pt} 
         \setlength{\itemsep}{0pt} \settowidth
	{\labelwidth}{#1.}\sloppy}}{\end{list}}

\newcounter{itemlistc}
\newcounter{romanlistc}
\newcounter{alphlistc}
\newcounter{arabiclistc}

\def\@citex[#1]#2{\if@filesw\immediate\write\@auxout
	{\string\citation{#2}}\fi
\def\@citea{}\@cite{\@for\@citeb:=#2\do
	{\@citea\def\@citea{,}\@ifundefined
	{b@\@citeb}{{\bf ?}\@warning
	{Citation `\@citeb' on page \thepage \space undefined}}
	{\csname b@\@citeb\endcsname}}}{#1}}

\newif\if@cghi
\def\cite{\@cghitrue\@ifnextchar [{\@tempswatrue
	\@citex}{\@tempswafalse\@citex[]}}
\def\citelow{\@cghifalse\@ifnextchar [{\@tempswatrue
	\@citex}{\@tempswafalse\@citex[]}}
\def\@cite#1#2{{$\null^{#1}$\if@tempswa\typeout
	{IJCGA warning: optional citation argument 
	ignored: `#2'} \fi}}

\def\@refcitex[#1]#2{\if@filesw\immediate\write\@auxout
	{\string\citation{#2}}\fi
\def\@citea{}\@refcite{\@for\@citeb:=#2\do
	{\@citea\def\@citea{, }\@ifundefined
	{b@\@citeb}{{\bf ?}\@warning
	{Citation `\@citeb' on page \thepage \space undefined}}
	\hbox{\csname b@\@citeb\endcsname}}}{#1}}

\def\@refcite#1#2{{#1\if@tempswa\typeout
        {IJCGA warning: optional citation argument
	ignored: `#2'} \fi}}

\def\refcite{\@ifnextchar[{\@tempswatrue
	\@refcitex}{\@tempswafalse\@refcitex[]}}

\def\pmb#1{\setbox0=\hbox{#1}
	\kern-.025em\copy0\kern-\wd0
	\kern.05em\copy0\kern-\wd0
	\kern-.025em\raise.0433em\box0}

\def\fnt#1#2{\footnotetext{\kern-.3em
	{$^{\mbox{\scriptsize #1}}$}{#2}}}

\def\runninghead#1#2{\pagestyle{myheadings}
\markboth{{\protect\footnotesize\it{\quad #1}}\hfill}
{\hfill{\protect\footnotesize\it{#2\quad}}}}
\font\tenrm=cmr10
\font\tenit=cmti10 
\font\tenbf=cmbx10
\font\bfit=cmbxti10 at 10pt
\font\ninerm=cmr9

\font\eightrm=cmr8

\def\qed{\hbox{${\vcenter{\vbox{			
   \hrule height 0.4pt\hbox{\vrule width 0.4pt height 6pt
   \kern5pt\vrule width 0.4pt}\hrule height 0.4pt}}}$}}

\begin{document}

\newpage

\runninghead{Rosu \& L\'opez-Sandoval}
{FRW barotropic cosmologies with Dirac}

\normalsize\textlineskip
\thispagestyle{empty}
\setcounter{page}{1}


\vspace*{0.88truein}

\bigskip
\centerline{\bf  BAROTROPIC FRW COSMOLOGIES WITH A DIRAC-LIKE PARAMETER} 
\vspace*{0.035truein}
\vspace*{0.37truein}
\vspace*{10pt}
\centerline{\footnotesize H.C. ROSU\footnote{E-mail: hcr@ipicyt.edu.mx; file romancos.tex} $\,$ 
and R. L\'OPEZ-SANDOVAL\footnote{E-mail: sandov@ipicyt.edu.mx}}
\vspace*{0.015truein}
\centerline{\footnotesize Potosinian Institute of Science and Technology,}
\centerline{ \footnotesize IPICyT, Apdo Postal 3-74 Tangamanga, 78231 San Luis Potos\'{\i}, Mexico}
\vspace*{0.225truein}


\vspace*{0.21truein}
\abstracts{Using the known connection between Schroedinger-like equations and Dirac-like equations in the 
supersymmetric context, we discuss an extension of FRW barotropic cosmologies in which a Dirac mass-like parameter is introduced.
New Hubble cosmological parameters $\rm H_K(\eta)$ depending on the Dirac-like parameter are plotted and compared with the standard Hubble case $\rm H_0(\eta)$. 
The new $\rm H_K(\eta)$ are complex quantities. The imaginary part is a supersymmetric way of introducing dissipation and instabilities in the 
barotropic FRW hydrodynamics.  
}{}{}


\textlineskip                  
\vspace*{12pt}                 

\vspace*{1pt}\textlineskip	
\vspace*{-0.5pt}
\noindent


\noindent

\bigskip





{\bf 1} - {\bf Introduction}
\medskip

\noindent
The scale factor ${\rm a(t)}$ of a FRW metric is a function of the
comoving time ${\rm t}$ obeying the Einstein-Friedmann 
dynamical equations supplemented by the (barotropic) equation of state of the cosmological fluid
\begin{equation} \label{e1}
{\rm \frac{\ddot{a}}{a}=-\frac{4\pi G}{3}(\rho +3p)}~,
\end{equation}
\begin{equation} \label{e2}
{\rm
\left(\frac{\dot{a}}{a}\right)^2=\frac{8\pi G\rho}{3}-\frac{\kappa}{a^2}~,}
\end{equation}
\begin{equation} \label{e3}
{\rm p=(\gamma -1)\rho,}
\end{equation}
where $\rho$ and ${\rm p}$ are the energy
density and the pressure, respectively,
of the perfect fluid of which a classical universe is usually assumed to be
made of, $\kappa=0,\pm1$ is the curvature index of the flat, closed, open
universe, respectively, and $\gamma$ is the constant adiabatic index of the cosmological fluid.
A simple {\em Riccati route} of solving the system of eqs. (1)-(3) proposed by Faraoni,\cite{Far} has been used by Rosu to develop
a factorization (nonrelativistic supersymmetry) approach of the barotropic FRW cosmologies,\cite{rosu} in which a simple explanation 
for a currently accelerating universe,\cite{type1a} in terms of `Darboux fluids' with effective adiabatic indices is possible.

In this work, we first briefly review the supersymmetric factorization methods for barotropic FRW cosmologies in section 2.
Next, in section 3, we present Dirac-like (first-order) coupled differential equations and their associated 
second-order differential equations. 
This allows a simple extension of barotropic FRW cosmologies that include a Dirac {\em mass-like} constant parameter ${\rm K}$.


\bigskip
{\bf 2} - {\bf Supersymmetric methods}

\medskip

\noindent
Combining the equations (1)-(3) and using the conformal time variable $\eta$  defined by
${\rm dt=a(\eta)d\eta}$ one gets the equation
\begin{equation} \label{comb}
{\rm \frac{a^{''}}{a}+(c-1)\left(\frac{a^{'}}{a}\right)^2+c\kappa=0~.}
\end{equation}
where ${\rm c=\frac{3}{2}\gamma -1}$. The case $\kappa=0$ is directly
integrable,\cite{Far} and will be skipped henceforth.
The logarithmic derivative
${\rm H_0=\frac{a^{'}}{a}}$ is the Hubble parameter of the universe which is a fundamental cosmological quantity ever since Hubble's discovery of the expansion of the 
universe in1929.
Interestingly, passing to the function ${\rm H_0}$ in Eq.~(\ref{comb}) 
the following Riccati equation is obtained
\begin{equation} \label{ricc}
{\rm H^{'}_{0}+cH^2_{0}+\kappa c=0~.}
\end{equation}
Employing now ${\rm H_0=\frac{1}{c}\frac{w^{'}_{\kappa}}{w_{\kappa}}}$ one gets the
very simple second order differential equation
\begin{equation} \label{schr1}
{\rm w^{''}_{\kappa}+\kappa c^2w_{\kappa}=0~.}
\end{equation}
For $\kappa =1$ the solution of the latter is
${\rm w_{1}=W_1\cos(c\eta +d)}$ , where ${\rm d}$ is an arbitrary phase,
implying
$
{\rm a_{1}(\eta)=A_1[\cos(c\eta +d)]^{1/c}~,}
$
whereas for $\kappa =-1$ one gets
${\rm w_{-1}=W_{-1}{\rm sinh}(c\eta)}$ and therefore
$
{\rm a_{-1}(\eta)=A_{-1}[{\rm sinh}(c\eta)]^{1/c}}~,
$
where ${\rm W_{\pm 1}}$ and ${\rm A_{\pm 1}}$ are amplitude parameters.


The point now is that the particular Riccati solutions
${\rm H_{0}^{+}=-\tan c\eta}$ and ${\rm H_{0}^{-}={\rm coth} c\eta}$
for $\kappa =\pm 1$, respectively, are
closely related to the common factorizations of the Schr\"odinger equation.
Indeed, Eq.~(\ref{schr1}) can be written
\begin{equation} \label{w}
{\rm w^{''}-c(-\kappa c)w=0}
\end{equation}
and also in factorized form using Eq.~(6) one gets
\begin{equation} \label{w3}
{\rm \left(\frac{d}{d\eta}+cH_{0}\right)
\left(\frac{d}{d\eta}-cH_{0}\right)w=
w^{''}-c(H_{0}^{'}+cH_{0}^{2})w=0}~.
\end{equation}
To fix the ideas, we shall call Eq.~(\ref{w3}) the bosonic equation.
On the other hand, the supersymmetric partner (or fermionic)
equation of Eq.~(\ref{w3}) will be
\begin{equation} \label{f}
{\rm
\left(\frac{d}{d\eta}-cH_{0}\right)
\left(\frac{d}{d\eta}+cH_{0}\right)w_f=
w^{''}_{f}-c(-H_{0}^{'}+cH_{0}^2)w_{f}={\rm w^{''}_f
-c\cdot c_{\kappa, susy}w_f=0}}~.
\end{equation}
Thus, one can write
$$
{\rm c_{\kappa,susy}(\eta)=-H_{0}^{'}+cH_{0}^2=
\left\{ \begin{array}{ll}
c(1+2{\rm tan}^2 c\eta) & \mbox{if $\kappa =1$}\\
c(-1+2{\rm coth}^2 c\eta) & \mbox{if $\kappa =-1$}
\end{array} \right.}
$$
for the supersymmetric partner adiabatic index.
The solutions are $\rm w_f =\frac{c}{\cos (c\eta +d)}$ 
and $\rm w_f =\frac{c}{sinh (c\eta)}$ for $\kappa =1$ and $\kappa =-1$,
respectively.

\bigskip
{\bf 3} -  {\bf Dirac-like formalism}

\medskip

\noindent
The Dirac equation in the susy nonrelativistic formalism has been discussed by Cooper {\em et al},\cite{cooper} already in 1988.
They showed that the Dirac equation with a Lorentz scalar potential is associated with a susy pair of Schroedinger Hamiltonians.
This result has been used later by many other authors. Here we make an application to barotropic FRW cosmologies that we find 
not to be a trivial exercise except for the uncoupled `zero-mass' case (subsection {\bf 3.1}). 

{\bf 3.1}-
Let's introduce now the following two Pauli matrices 
$$\alpha =-i\sigma _y=-i\left( \begin{array}{cc}
0 & -i \\
i & 0\end{array} \right ) \qquad {\rm and} \quad
\beta =\sigma _x=\left( \begin{array}{cc}
0 & 1\\
1 & 0 \end{array} \right )
$$
and write a cosmological Dirac equation
\begin{equation} \label{HD}
{\rm {\cal H}_{0}^{{\rm FRW}}W=[\sigma _y D_{\eta}+\sigma _x (icH_0)]W=0}~,
\end{equation}
where $W=\left( {\rm \begin{array}{cc}
w_1\\
w_2\end{array}} \right ) $ is a two component `zero-mass' spinor. 
This is equivalent to the following decoupled equations
\begin{eqnarray}
{\rm D}_{\eta}w_1+{\rm cH_0}w_1=0 \\                                     
-{\rm D}_{\eta}w_2+{\rm cH_0}w_2=0~.                                                                            
\end{eqnarray}
Solving these equations one gets $w_1\propto 1/\cos ({\rm c}\eta)$ and $w_2\propto \cos({\rm c}\eta)$ for $\kappa =1$ cosmologies
and $w_1\propto 1/{\rm sinh} ({\rm c}\eta)$ and $w_2\propto {\rm sinh}({\rm c}\eta)$ for $\kappa =-1$ cosmologies.
Thus, we obtain 
$$
W=\left( {\rm \begin{array}{cc}
w_1\\
w_2\end{array}} \right )=\left( {\rm \begin{array}{cc}
{\rm w_f}\\
{\rm w_b}\end{array}} \right )~.
$$
This shows that the matrix `zero-mass' Dirac equation is equivalent to the two Schroedinger equations for the bosonic and fermionic 
components.  

\bigskip

{\bf 3.2}-
Consider now a ``massive" Dirac equation 
\begin{equation} \label{HDM}
{\rm H_{K}^{FRW}W=[\sigma _y D_{\eta}+\sigma _x (icH_p +K)]W=KW}~,
\end{equation}
where $\rm K$ may be considered the mass parameter of the Dirac spinor. Eq.~(\ref{HDM}) 
is equivalent to the following system of coupled equations
\begin{eqnarray}
{\rm D_{\eta}w _1+(i{\rm cH_0+K})w _1={\rm K}w _2}\\
-{\rm D_{\eta}w _2+({\rm icH_0+K})w _2={\rm K}w _1}~.
\end{eqnarray}
These two coupled first-order equations are equivalent to second order differential equations for the two spinor components:

\noindent
The fermionic spinor component can be found directly as solutions of
\begin{equation} \label{comp1} 
{\rm D^{2}_{\eta}w_1^{+}-\Big[c^2(1+2\tan ^2 c\eta)+2icK\tan c\eta\Big] w_1^{+}=0}  \qquad {\rm for} \, \kappa =1
\end{equation}
and
\begin{equation} \label{comp1b} 
{\rm D^{2}_{\eta}w_1^{-}-\Big[c^2(-1+2{\rm coth}^2 c \eta)-2icK{\rm coth} \,c\eta\Big] w_1^{-}=0}  \qquad  {\rm for} \, \kappa =-1~,
\end{equation}
whereas the bosonic components are solutions of
\begin{equation} \label{comp2} 
{\rm D^{2}_{\eta}w_2^{+}+\Big[c^2-2icK\tan c\eta\Big] w_2^{+}=0}  \qquad {\rm for} \quad  \kappa =1
\end{equation}
and 
\begin{equation} \label{comp2b} 
{\rm D^{2}_{\eta}w_2^{-}+\Big[-c^2+2icK{\rm coth} \,c\eta\Big] w_2^{-}=0}  \qquad {\rm for} \quad \kappa =-1~.
\end{equation}
The solutions of the bosonic equations are expressed in terms of the Gauss hypergeometric functions ${\rm _2F_1}$ in the variables ${\rm y=e^{ic\eta}}$ and
${\rm y=e^{c\eta}}$, respectively
$$
{\rm w_2^{+}=Ay^{-p}\, _{2}F_1\Big[-\frac{1}{2}(p+iq);-\frac{1}{2}(p-iq), 1-p; -y^2\Big]}+
$$
\begin{equation} \label{s1}
{\rm By^{p} \, _2F_1\Big[\frac{1}{2}(p-iq), \frac{1}{2}(p+iq),1+p; -y^2\Big]}
\end{equation}
and 
$$
{\rm w_2^{-}=C(-1)^{-\frac{i}{2}r}y^{-ir}\, _{2}F_1\Big[-\frac{i}{2}(r+s),-\frac{i}{2}(r-s), 1-r; y^2\Big]}+ 
$$
\begin{equation} \label{s2}
{\rm D(-1)^{\frac{i}{2}r}y^{ir}\, _{2}F_1\Big[\frac{i}{2}(r-s),\frac{i}{2}(r+s), 1+r; y^2\Big]}~,
\end{equation}
respectively. The parameters are the following: 
$$
{\rm p=(-1-\frac{2K}{c})^{\frac{1}{2}},  \quad q=(1-\frac{2K}{c})^{\frac{1}{2}}, \quad
r=(-1-i\frac{2K}{c})^{\frac{1}{2}}, \quad s=(-1+i\frac{2K}{c})^{\frac{1}{2}}},
$$ 
whereas ${\rm A}$, ${\rm B}$, ${\rm C}$, ${\rm D}$ are superposition constants.

\noindent
Based on these ${\rm K}$- modes, we can introduce bosonic Hubble parameters depending on the parameter ${\rm K}$
\begin{equation}\label{hw+}
{\rm H^{+}_{K}=\frac{1}{ c}\left(\frac{D}{D\eta}\log w_2^{+}\right)}
\end{equation}
and
\begin{equation}\label{hw-}
{\rm H^{-}_{K}=\frac{1}{c}\left(\frac{D}{D\eta}\log w_2^{-}\right)}~,
\end{equation}
and similarly for the fermionic components by changing ${\rm w_2^{\pm}}$ to ${\rm w_1^{\pm}}$ in eqs.~(\ref{hw+}) and (\ref{hw-}), respectively.

In the small ${\rm K}$ limit, ${\rm K\rightarrow 0}$, the ordinary FRW barotropic cosmologies are obtained. The Hubble 
parameters corresponding to the $\rm K$- dependent bosonic modes are plotted in the Figures~(1) -(4).

\bigskip
{\bf 4 - Conclusions}

\noindent
We come now to the interpretation of the mathematical results that we displayed in 
the previous sections.
The parameter $\rm K$ introduces an imaginary part in the cosmological Hubble parameter $\rm H$. Since the latter is the logarithmic derivative 
of the scale factor of the universe one comes to the conclusion that the supersymmetric techniques presented here are a way to
consider dissipation and instabilities of barotropic FRW cosmologies.





\begin{figure}[htb]
\centerline{
\includegraphics[scale=1]{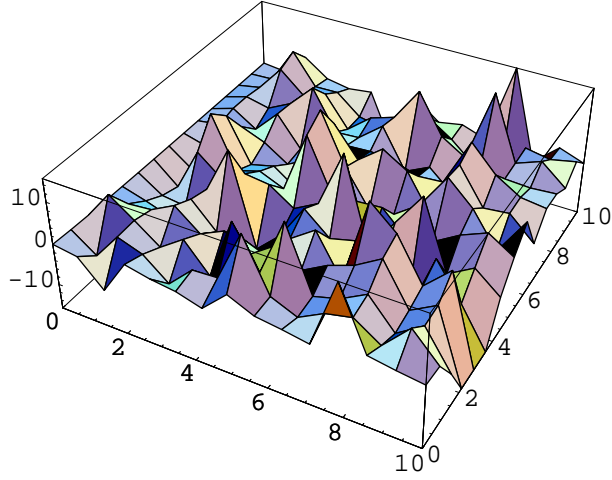}}
\caption{The real part of ${\rm H^{+}_{K}}$ obtained from the logderivative of the bosonic mode ${\rm w_2^{+}(y;\frac{1}{2},\frac{1}{2}})$ for ${\rm \eta}\in [0,10]$ and ${\rm K\in[0,10]}$ in the case of a radiation dominated universe (${\rm c=1}$) .
} \label{fig1ho}
\end{figure}

\begin{figure}[htb]
\centerline{
\includegraphics[scale=1]{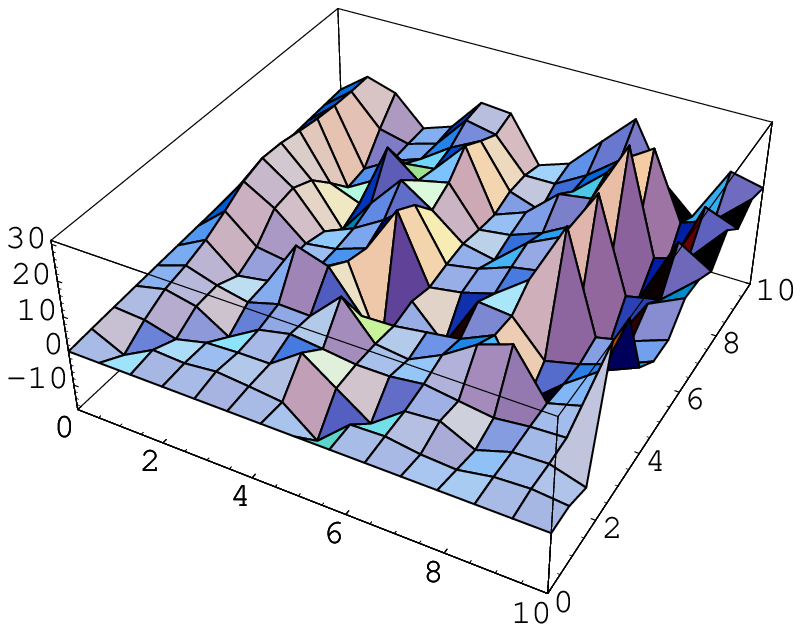}}
\caption{The imaginary part of ${\rm H^{+}_{K}}$ for the same case.  
} \label{fig2ho}
\end{figure}

\begin{figure}[htb]
\centerline{
\includegraphics[scale=1]{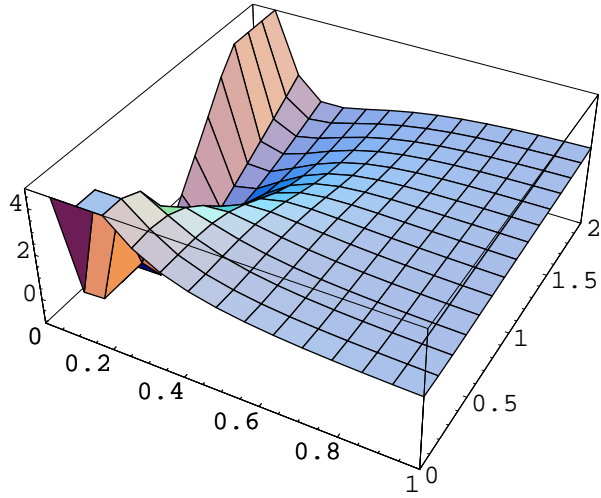}}
\caption{The real part of  ${\rm H^{-}_{K}}$ calculated from the bosonic mode ${\rm w_2^{-}(y;\frac{1}{2},\frac{1}{2}})$ for ${\rm \eta}\in [0,1]$ and ${\rm K\in[0,2]}$
and a radiation dominated universe.
} \label{fig3ho}
\end{figure}

\begin{figure}[htb]
\centerline{
\includegraphics[scale=1]{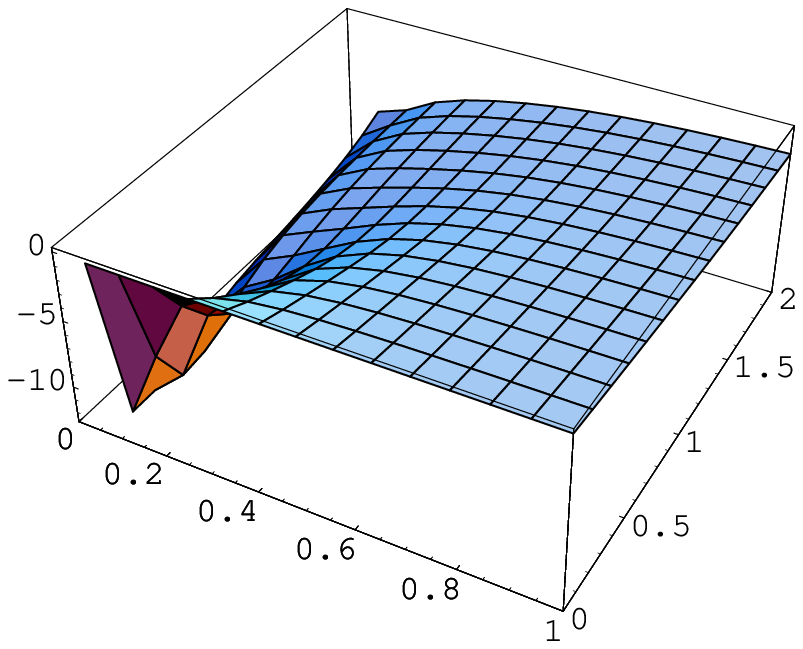}}
\caption{The imaginary part of ${\rm H^{-}_{K}}$ for the same case. 
} \label{fig4ho}
\end{figure}

\end{document}